\documentclass{iaus}
\usepackage{graphicx}
\usepackage{natbib}
\usepackage{rotating}

\newcommand\apj{{ApJ}}
\newcommand\apjl{{ApJ}}
\newcommand\aap{{A\&A}}
\newcommand\mnras{{MNRAS}}
\newcommand\nat{{Nature}}
\newcommand\icarus{{Icarus}}

\title[Planet--planet scattering] 
{How planet--planet scattering can create high-inclination as well as long-period orbits}

\author[Chatterjee\ et\ al.]   
{Sourav Chatterjee$^1$
 , Eric B. Ford$^1$, \and Frederic A. Rasio$^2$}
\affiliation{$^1$University of Florida, 211 Bryant Space Science Center, Florida, USA \\
email: {\tt s.chatterjee@astro.ufl.edu}, {\tt eford@astro.ufl.edu} \\[\affilskip]
$^2$CIERA, Northwestern University, Evanston, IL 60208, USA \\email: {\tt rasio@northwestern.edu}}

\pubyear{2010}
\volume{276}  
\pagerange{119--126}
\jname{Observations and theory of exoplanets}
\editors{A. Sozzetti, M. G. Lattanzi, B.D. Editor \& A. P. Boss, eds.}
\begin{document}

\maketitle
\begin{abstract}
Recent observations have revealed two new classes of planetary orbits.  
Rossiter-Mclaughlin (RM) measurements have revealed hot Jupiters in high-obliquity 
orbits.  In addition, direct-imaging has discovered giant planets at large ($\sim 100\,\rm{AU}$) 
separations via direct-imaging technique.  Simple-minded disk-migration scenarios 
are inconsistent with the high-inclination (and even retrograde) orbits as seen in recent 
RM measurements.  Furthermore, 
forming giant planets at large semi-major axis ($a$) may be challenging in the core-accretion 
paradigm.  We perform 
many $N$-body simulations to explore the two above-mentioned orbital architectures. 
Planet--planet scattering in a multi-planet system can naturally excite orbital inclinations.  
Planets can also get scattered to large distances.  
Large-$a$ planetary orbits created from planet--planet scattering are expected to 
have high eccentricities ($e$).     
Theoretical models predict that the observed long-period planets, such as Fomalhaut-b 
have moderate $e\approx 0.3$.  Interestingly, these 
are also in systems with disks.  We find that if a massive-enough outer disk is present, a scattered 
planet may be circularized at large $a$ via dynamical friction from the disk and 
repeated scattering of the disk particles.   
\end{abstract}
\firstsection 
\section{Introduction}
The $15$ years since the discovery of the first exoplanet around a Solar-like star 
\citep{1995Natur.378..355M} have seen a revolution in our understanding of planet formation 
and evolution.  Observations and theoretical modeling have worked 
hand-in-hand to discover and explain new architectures of planetary orbits.  It is now 
well known that many exoplanets have large $e$ compared to our 
Solar system planetary orbits, indicating an active dynamical history 
\citep[e.g.,][]{Chatterjee08,Juric08,Nagasawa08}.  

Recent RM measurements of many transiting planets are putting 
further constraints on theoretical models of various planet formation and evolution scenarios 
\citep[e.g.,][]{2010arXiv1008.2353T,2010ApJ...718L.145W,2010arXiv1010.4025M}.  
Indeed, recent measurements find a large population of highly inclined planetary 
orbits, and even some retrograde orbits (see contributions from 
Winn et\ al. and Triaud et\ al. in this volume).  
Disk--planet migration models generally predict alignment between the planetary orbital angular momentum 
and the stellar spin axis from an aligned protoplanetary disk.    
Thus high-inclination orbits suggest dynamical evolution to 
be important in shaping the exoplanet architectures.  Alternatively, inclined orbits might also 
result if the inner portion of the protoplanetary disk itself had been misaligned 
\citep[][see also Lai et\ al. in this volume]{2010arXiv1008.3148L}.    

Recent high-contrast imaging has 
revealed another class of planets---giant planets at very large 
$a$ \citep[$\gtrsim50\,\rm{AU}$; e.g.,][]{Fomb_Kalas,HR8799_Marois,
2008A&A...477L...1D,2010arXiv1011.2201I}.  Timescale considerations for the core-accretion 
model of planet formation indicates that forming these planets in situ may be hard 
\citep[e.g.,][]{2001Icar..153..224L,2009ApJ...707...79D}.  Instead, we consider formation 
at closer orbital separations followed by planet--planet 
scattering to launch planets in large-$a$ orbits from dynamically active 
systems \citep[e.g.][]{Chatterjee08,Juric08,Nagasawa08}.  However, these simulations predict 
that these orbits typically also have high $e\gtrsim0.6$.  
Interestingly, some of the observed large-$a$ systems also have disks 
\citep[e.g.][]{Fomb_Kalas} and dynamical modeling of these disks 
indicates moderate values of $e\approx0.3$.  
We have started to explore the possibility that a planet launched into 
a large-$a$ and high-$e$ orbit can later be circularized near its apocenter if the planet 
enters a debris disk during its apocenter excursion.  

In Section\ \ref{inclination} we summarize our numerical setup and present results for 
expected inclinations from planet--planet scattering.  In 
Section\ \ref{long} we discuss how planet--planet scattering followed by circularization 
due to a residual disk may create moderate-$e$ giant planets at large $a$.  In 
Section\ \ref{conclude} we conclude.          
\section{Orbital inclinations from planet--planet scattering}
\label{inclination}
\begin{figure}[h]
\begin{center}
$\begin{array}{cc}
 \includegraphics[width=2.6in]{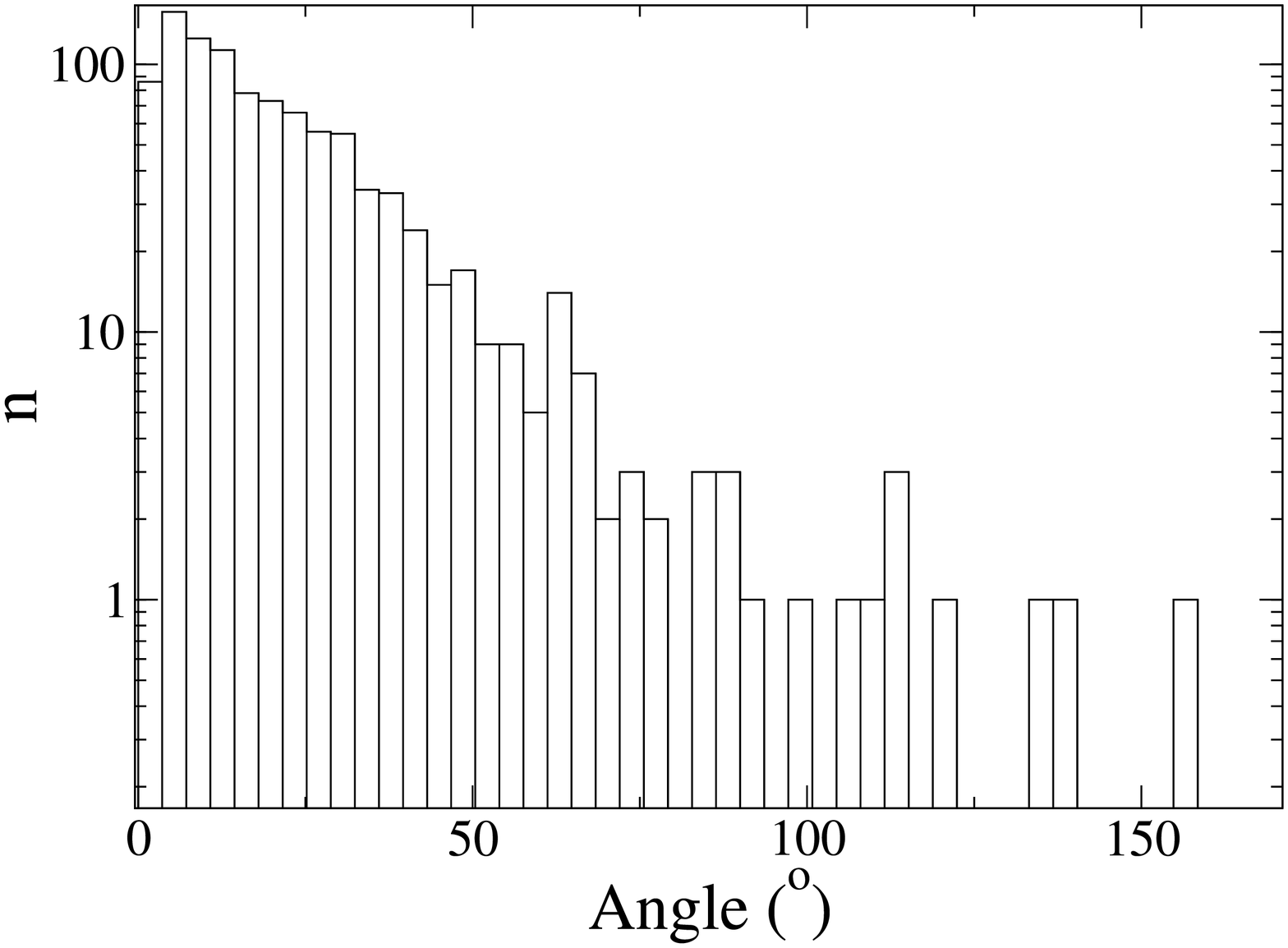} & 
 \includegraphics[width=1.9in]{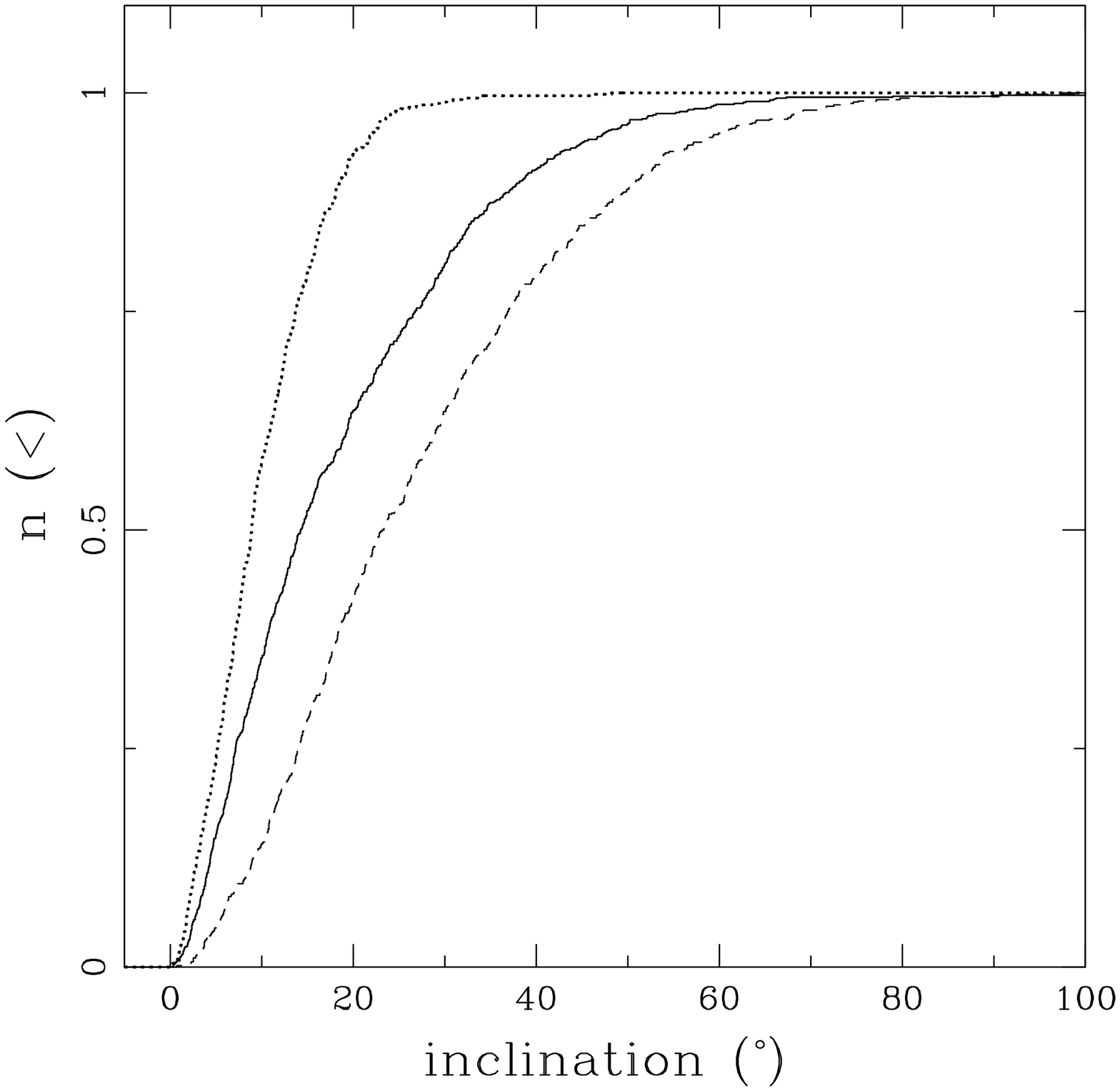} 
 \end{array}$
 \caption{Distribution of orbital inclinations with respect to the initial invariable plane.  
 {\it Left:} Histogram for the final inner-planet orbit.  In our simulations about $2\%$ of final 
 inner planets are in retrograde orbits.  
 {\it Right:} Cumulative histograms for the final inner (solid), and outer (dotted) planets.  
  The relative inclinations between the inner and outer orbits (dashed) are also shown.  About 
  $20\%$ of our simulated systems with two giant planets at the end have 
  relative inclination angles $\geq 40\,^\circ$ and could later go through Kozai-type oscillations.  
}
   \label{fig:inclination}
\end{center}
\end{figure}
We have simulated $3$ giant planets with masses between $0.4$--$1.2$ Jupitar-mass 
($\rm{M_J}$) around a Solar mass star.  The initial innermost planet is placed at $3\,\rm{AU}$, 
and the other $2$ planets are placed with planet--planet separation of $4.4$ Hill radii.  The initial 
$a$'s were chosen to avoid Mean-motion 
resonances.  The initial $e$ for the planetary orbits 
are chosen uniformly between $0$--$0.1$.  Initial $i$ is chosen uniformly between 
$0\,^\circ$--$10\,^\circ$ with respect to the initial innermost orbital plane.  All phase angles are 
assigned random values in the full range.  Each of the above configurations is integrated using 
the hybrid integrator of MERCURY6.2 \citep{Chambers_Mercury} for $10^7\,\rm{yr}$.  If the 
energy conservation is poorer than $10^{-3}$ then the full integration is repeated 
using the Burlisch-Stoeer (B-S) integrator (see \citealt{Chatterjee08} for more details).  

Figure\ \ref{fig:inclination} shows the distributions of the final inner- and outer-planet 
orbital inclinations with respect to the initial invariable plane, as well as the relative 
inclinations between the orbits.  
We find that planet--planet scattering is very efficient at exciting the inclination 
of planetary orbits.  These high inclinations are not mere reflection of the initial inclinations.  
There is in fact no correlation between the initial and final inclinations.  Starting from only moderate 
inclinations planet--planet scattering can create large inclinations extending all the way to 
retrograde orbits (although only about $2\%$ in these simulations).  
The relative inclinations between the planetary orbits in systems where 
two giant planets remain bound are also high (median $\approx 25\,^\circ$).  Note that 
the median value is incidentally in good agreement with the recent relative inclination 
measurement for $\nu-$Andromidae \citep[c \& d, $29.9\,^\circ\pm1\,^\circ$;][]{2010ApJ...715.1203M}.  
In our simulations about $20\%$ systems 
show relative inclinations angles $\geq 40\,^\circ$, making them potentially Kozai active.                    
\section{Long-period planets}
\label{long}
\begin{figure}[h]
\begin{center}
$\begin{array}{cc}
 \includegraphics[width=2.5in]{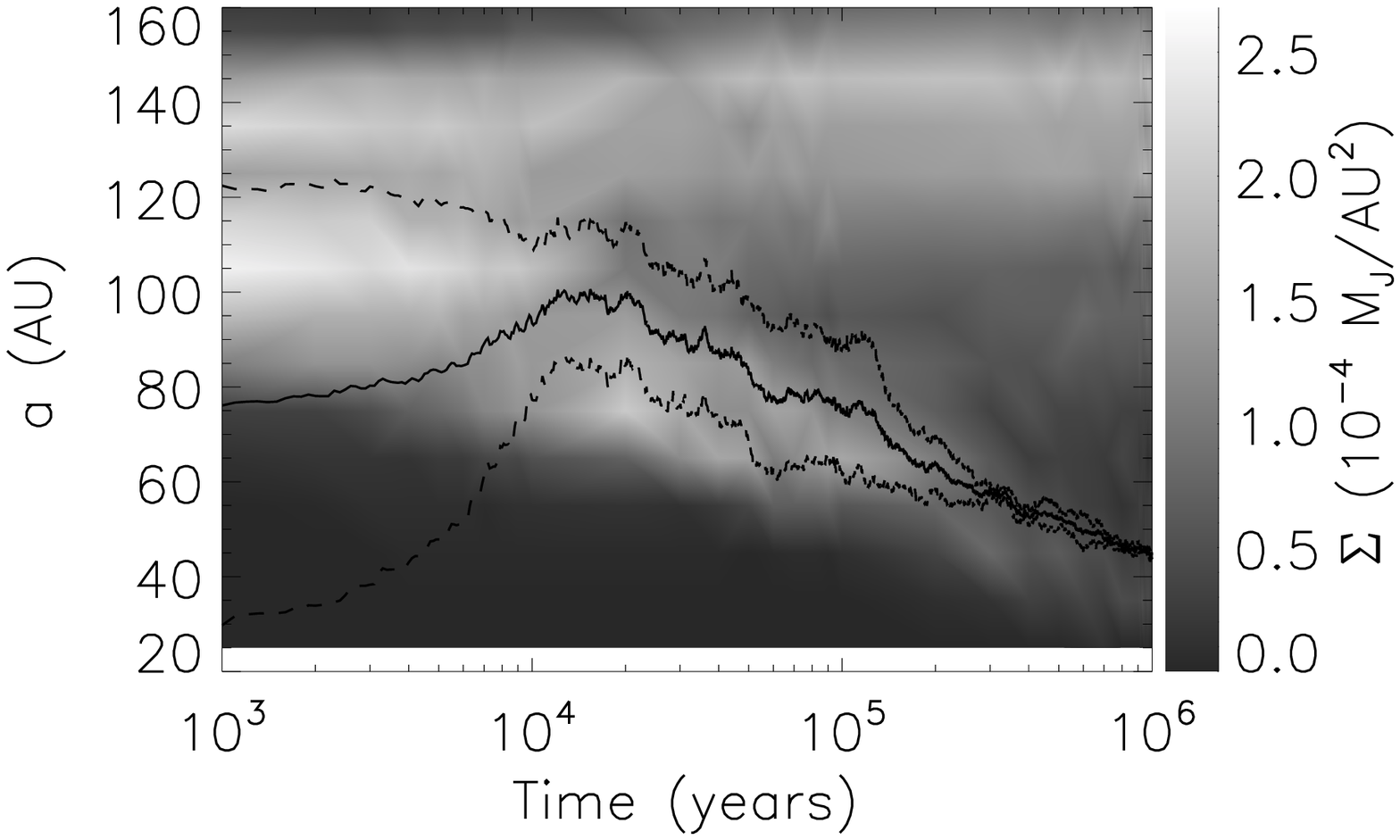} & 
 \includegraphics[width=2.5in]{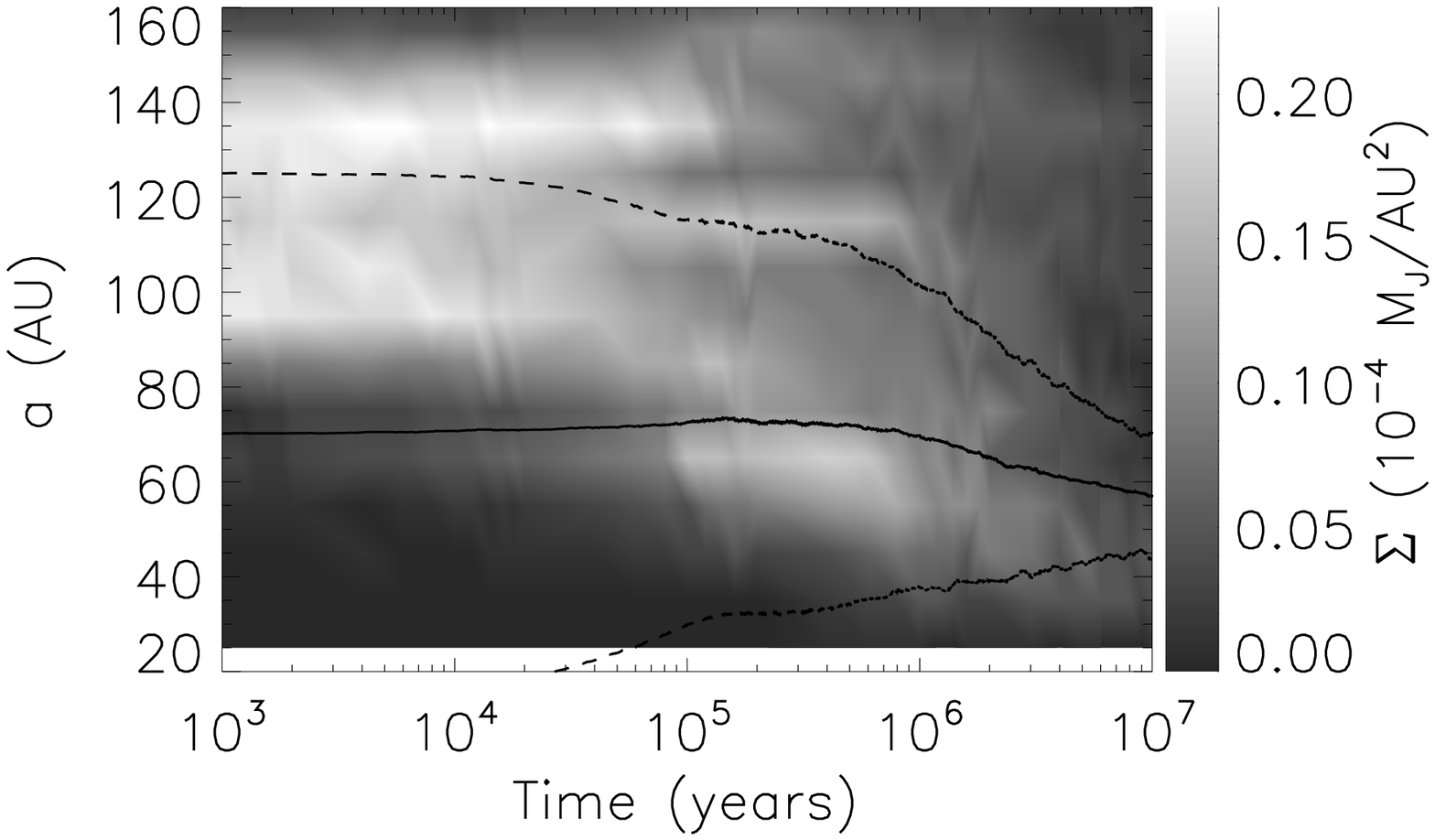}
 \end{array}$
 \caption{Evolution of the planetary orbit and surface density of the disk.  {\it Left:} Evolution for 
 the higher density disk {\tt model1}.  {\it Right:} Evolution for the lower density disk {\tt model2}.  
 The planets migrate outwards at first due to dynamical friction from the disk.  The 
 later inward migration of the planets is due to random outward scattering events of the disk 
 particles by the planet.  Planetary $e$ decreases throughout the entire evolution.    
 }
   \label{fig:long}
\end{center}
\end{figure}
Planet--planet scattering naturally creates large-$a$ orbits.  
In our simulations with $3$ giant planets we find outer planets with 
$a$ up to about $150\,\rm{AU}$.  
Here we explore the following scenario.  A massive disk of rocky material 
(possibly relic from the planet-formation process) remains at large separation from the star.  
This disk can remain relatively undisturbed for some time while the planets 
remain much closer to the star.  At some point in the evolution, planet--planet scattering 
launches one of the planets into a large-$a$, and high-$e$ orbit so that the giant planet enters the 
disk near apocenter.  The Keplerian velocity 
of the planet near apocenter is less than that of the material in the disk as long as the $e$ of the 
planet is higher than the typical $e$ of the disk material.  Thus, while in the disk the planet 
experiences a force due to dynamical friction from the disk in the direction of the 
planet's orbital velocity, increasing the planet's orbital energy and angular momentum.  
As a result, the planet's pericenter is raised as the orbit is circularized.  
The amount of migration and the timescale for circularization are directly proportional to the 
disk density.  

Using the B-S integrator in MERCURY6.2 we simulate the evolution of a giant planet orbit 
initially in a large-$a$ and high-$e$ orbit (possibly created via a planet--planet scattering 
event).  The initial $a$, and $e$ of the planetary orbit are $70\,\rm{AU}$, and $0.7$, respectively.  
Note that, in this case our $t = 0$ is after a planet--planet scattering event in a multi-giant 
planetary system that has launched a giant planet into this orbit.  
A disk of material is distributed initially between 
$90$--$150\,\rm{AU}$ with a uniform surface density.  The mass in the disk is represented in our 
simulations by $10^3$ equal-mass particles.  The disk particles interact with the planet but 
not with each other.  Initial $e$ and inclination of the disk material are chosen 
uniformly between $0$--$0.3$ and $0\,^\circ$--$4\,^\circ$, respectively.  The planetary 
orbit is assumed to be aligned with the mid-plane of the disk (as a first step).  
We use two models varying the initial disk surface densities ($\Sigma$) keeping everything 
else fixed.  Our {\tt model1} and {\tt model2} have $\Sigma=10^{-4}$ and 
$10^{-5}\,\rm{M_JAU^{-2}}$, respectively.      

Figure\ \ref{fig:long} shows the evolution of the planetary orbit as well as the surface density 
contours of the disk.  For {\tt model1} (Figure\ \ref{fig:long}, left), 
the planet first migrates outwards 
from $70\,\rm{AU}$ to $\approx 100\,\rm{AU}$ in $\approx 10^4\,\rm{yr}$.  Note that the outward 
migration happens via circularization of the planetary orbit near the planet's apocenter.  At the 
end of this migration the planet is in an orbit with $a=100\,\rm{AU}$ and $e=0.14$.  The outward 
migration is halted when the planet's intrusion severely depletes the disk.  Until then the 
planet's evolution is dominated by dynamical friction arising from the disk.  During this 
process the planet scatters a part of the disk material inwards (seen as a strip of over-density 
extending inwards).  The subsequent evolution of the planetary orbit is governed by random 
scattering events between the planet and disk particles.  
The planet migrates inwards by scattering outwards disk particles that the planet 
had previously scattered inwards.  This is a much slower process compared to the initial 
dynamical-friction dominated outward migration.  The $e$ is damped throughout the entire 
process.  At the integration stopping time ($10^6\,\rm{yr}$) the planet is in an orbit 
with $a=44\,\rm{AU}$ and $e=0.02$.  

Qualitatively similar behavior is noted in the evolution of the planetary orbit for {\tt model2}.  
However, in this case the timescale of the outward migration is about $10$ times 
longer ($\sim10^5\,\rm{yr}$) and the magnitude of migration is less by a 
factor of $10$ compared to {\tt model1}.  At the end of the 
dynamical-friction-driven outward migration the planetary orbit has $a=74\,\rm{AU}$ and 
$e=0.5$.  This stage of evolution 
is followed by the planet--disk particle scattering stage during which the planet migrates inwards.  
At the integration stopping time ($10^7\,\rm{yr}$) for {\tt model2} the planetary orbit 
has $a=57\,\rm{AU}$ and $e=0.2$.  

In both cases at the end the planet is (or will be) left in a large-$a$, and modest-$e$ orbit.  
In addition, 
the total mass in the disk is reduced dramatically.  The intrusion of the giant planet also 
excites the $e$ and inclination of the disk material.  Furthermore, a low density disk may remain 
outside the planetary orbit which may then, over time, grind to create a dust ring via collisions, 
similar to the one observed in Fomalhaut-b.                     
\section{Discussion}
\label{conclude}
Using numerical simulations we have studied how planet--planet scattering can naturally 
create high-inclination orbits.  We have studied whether giant planets can be launched into 
large-$a$, but modest-$e$ orbits, similar in architecture to Fomalhaut-b, via 
planet--planet scattering.  We find that planet--planet scattering can naturally create many 
large-$a$ orbits, however, these orbits are also expected to have high $e$ 
\citep[e.g.,][]{Chatterjee08}.  Nevertheless, if a massive outer disk exists 
and the scattered giant planet near its apocenter enters the disk, 
dynamical friction from the disk can raise the planet's $a$ until 
the planet reduces the disk density significantly via scattering.  Then the planet 
can migrate inwards via scattering some of the remaining disk particles outwards.  
During the whole process the $e$ of the planetary orbit reduces.  We plan to further study 
this process in detail exploring different disk masses, densities, as well as varying planet masses.  
We thank the SOC and LOC for arranging this excellent conference and the 
opportunity to present these results.      

\end{document}